# Large-scale pedestrian flow experiments under high-density conditions

Cheng-Jie Jin [1,2], Rui Jiang [3], S.C. Wong [4], Dawei Li [1,2], Ning Guo [5], Wei Wang [1,2]

[1] *Jiangsu Key Laboratory of Urban ITS, Southeast University of China, Nanjing, Jiangsu, 210096, People's Republic of China*

[2] *Jiangsu Province Collaborative Innovation Center of Modern Urban Traffic Technologies, Nanjing, Jiangsu, 210096, People's Republic of China*

[3] *MOE Key Laboratory for Urban Transportation Complex Systems Theory and Technology, Beijing Jiaotong University, Beijing, 100044, People's Republic of China*

[4] *Department of Civil Engineering, The University of Hong Kong, Hong Kong, People's Republic of China*

[5] *School of Engineering Science, University of Science and Technology of China, Hefei, Anhui, 230026, People's Republic of China*

**Abstract:**

Despite the vast amount of studies on pedestrian flow, the data concerning high densities are still very inadequate. We organize one large-scale pedestrian flow experiment on a ring corridor. With 278 participants, the density as high as 9 $m^{-2}$ is reached. In the uni-directional flow, four different states are observed, including the free flow, congested state, over-congested state and hyper-congested state. The features of the hyper-congested state are similar to the "crowd turbulence" reported in the empirical data of Helbing et al., and the transition between the stopped state and the moving state can be found. The flow rates in the over-congested state are nearly constant, due to the downstream propagation of pedestrian clusters. In the bi-directional flow, three different types of lane formations are observed in the experiment: (1) three lanes are directly formed ; (2) two lanes are directly formed; (3) firstly three lanes are formed, and then they transit into two lanes. After the lane formation, some interesting phenomena have been observed, including the inhomogeneous distribution of pedestrians across the lanes, and the formation and dissipation of localized crowd. Our study is expected to help for better understanding and modeling the dynamics of high density pedestrian flow.

**Keywords:** pedestrian flow; experiment; uni-directional flow; bi-directional flow; lane formation

## 1. Introduction

The study of pedestrian flow has a long history (Helbing and Molnár, 1995; Helbing, 2001; Hughes 2002; Hoogendoorn and Bovy, 2004; Hoogendoorn and Daamen, 2005; Huang et al., 2009; Papadimitriou et al., 2009; Wong et al., 2010; Flötteröd and Lämmel, 2015; Hoogendoorn et al., 2015; Ma and Yarlagadda, 2015; Nikolic et al., 2016). With the rapid development of cities and the growth of population, this study becomes more and more critical in recent days. The main purpose of pedestrian dynamics study is at least tri-fold: (i) to enhance the efficiency of the pedestrian facilities, such as various kinds of pedestrian passages; (ii) to guarantee the safety of pedestrians in high density crowd, in which the stampede accident is prone to occur; (iii) to increase the evacuation efficiency under panic situation due to fire, explosion and so on.

To study the features of pedestrian dynamics, many empirical and experimental data have been collected, and many simulation models have been proposed. The studies reveal the fundamental relationship between speed, density and flow rate, as well as some self-organization phenomena, such as the spontaneous lane formation in bidirectional flow, "fast is slow" effect and herding behavior in evacuation process, turbulent movement in the dense crowds, and alternation of passing directions at bottlenecks (Helbing et al., 2000, 2001, 2007; Seer et al., 2014).

This paper focuses on the empirical and experimental studies of pedestrian flow under non-panic situation. As shown in the literature review in Section 2, although there are many empirical and experimental studies of pedestrian flow, the data at high densities ($\rho > 6m^{-2}$) are still very inadequate. We would like to mention that stampede incidents usually happen at very high densities. Therefore, it is very necessary to investigate the pedestrian dynamics at these situations.

Motivated by the fact, we organized an experiment in a circular corridor. With 278 participants, the maximum density reaches $\rho = 9m^{-2}$. For uni-directional flow, the fundamental diagram has been obtained, and different states have been identified. For bi-directional flow, different types of lane formation processes have been observed and analyzed. Our study is expected to help for better understanding the dynamics of high density pedestrian flow.

The rest of this paper is organized as follows. Section 2 makes a literature review of the state-of-the-art of the empirical and experimental studies of pedestrian flow under non-panic condition. The configuration of the experiment is introduced in Section 3. The experimental results of uni-directional flow and bi-directional flow are discussed in Section 4 and 5, respectively. The conclusion is given in Section 6.

## 2. Literature review

This section makes a brief literature review for the empirical and experimental studies of pedestrian flow under non-panic condition, particularly focusing on the maximum density reported in the data.

### 2.1 Empirical studies

Fig.1 shows the empirical fundamental diagrams of several typical studies. Fig.1(a) shows the results of uni-directional flow. The Data of Helbing et al. (2007) were collected at the Hajj pilgrimage. The flow rate increases quickly and almost linearly when the density is below $1.5m^{-2}$. Then the flow rate increases slowly until the density reaches $5m^{-2}$ and suddenly drops. The flow rate remains low and almost constant when the density is above $6m^{-2}$. The maximum density observed reaches $10m^{-2}$. The data of Mori and Tsukaguchi (1987) were collected in Japan, from the commuters on the sidewalks. One can see that the flow rate is very close to that of Helbing et al. However, the maximum density observed only reaches $6m^{-2}$, and no drop of flow rate is observed. The uni-directional data of Oeding (1963) have two datasets: the first one was collected from the workers of different companies on footpaths of inner city mainroads; the second one was collected from workers moving to and coming from big factories. In the two datasets, the maximum density is no more than $3m^{-2}$.

Fig.1(b) shows the fundamental diagrams of bi-directional flow. The bi-directional data of Oeding (1963) also have two datasets: the first one was collected from inner city shopping streets, and the second one was obtained from pedestrians moving to and from a sports stadium. The data of Navin and Wheeler (1969) were collected on sidewalks at the University of Missouri and Stephens College. The data of Older (1968) were also collected on shopping streets. One can see that the maximum densities observed in all these datasets are smaller than $5m^{-2}$.

We would like to mention that there are many other empirical studies, see e.g., Lam et al. (2002, 2003), in which the bi-directional flows in indoor and outdoor walkways of Hong Kong have been studied. However, these studies only discuss the flow rates and the average velocities. The flow-density or velocity-density relationship has not been explicitly reported. There are also some studies concerning the high density crowd and some stampede accidents, see e.g., Zhang et al. (2012), Helbing and Mukerji (2012) and Krausz and Bauckhage (2012). In these studies, the video data are analyzed and discussed. But the flow-density or velocity-density relationship is not mentioned, either.

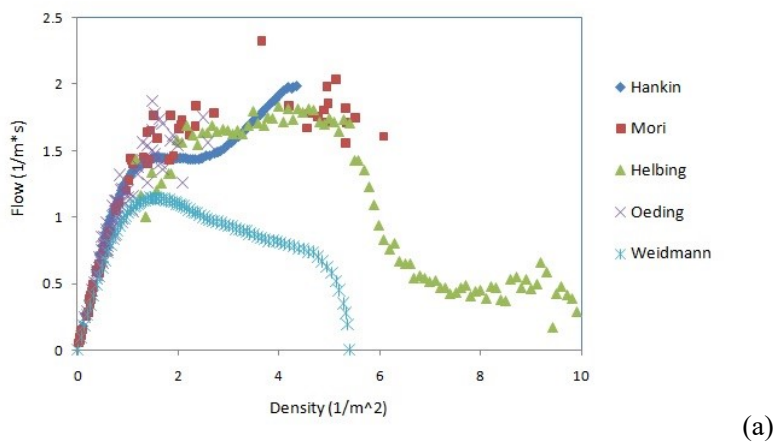

(a)

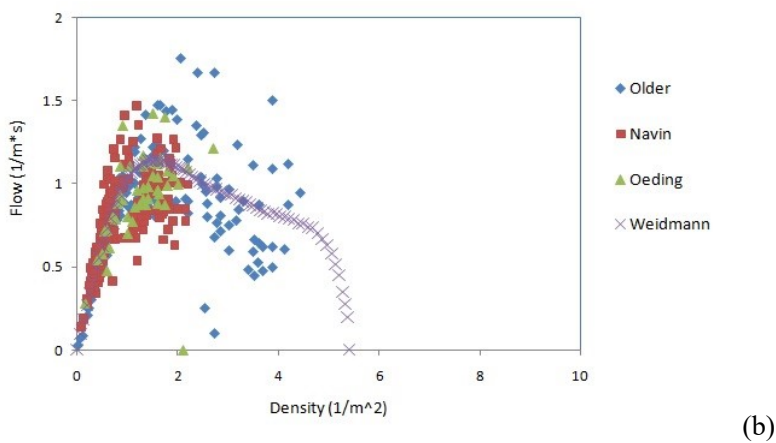

(b)

Fig.1. The fundamental diagrams of pedestrian flow. (a) uni-directional flow; (b) bi-directional flow. The data of Weidmann (1993) are used as a benchmark, which are the averaged results of 25 publications, including the uni- and bi-directional flow data observed in some field studies and experimental researches. All the datasets are downloaded at http://www.asim.uni-wuppertal.de/en/database/data-from-literature/fundamental-diagrams.html.

## 2.2 Experimental studies

2.2.1 Circular corridor

As early as in 1958, Hankin and Wright (1958) have performed an experimental study on uni-directional pedestrian flow in UK. The experiment track is a circular passageway with the width of 1.3m. The internal diameter is 9m. Over 200 boys were gradually fed into the ring. Thus, the maximum density can reach about 4.4 $m^{-2}$. See Fig.1(a) for the obtained flow-density relationship.

Moussaid et al. (2012) performed the experimental study on bi-directional pedestrian flow in France. The width of experiment track is 2.5m and the internal diameter is 2m. There are 30, 50 or 60 participants in different runs. The maximum density is thus only $1.2m^{-2}$. The experiment shows that the traffic segregation exhibits structural instabilities characterized by the alternation of organized and disorganized states.

Guo et al. (2016) performed an experiment in China to study the effect of limited view. The width of experiment track is 3m and the internal diameter is 4m. The number of participants is 100. Therefore, the maximum density is $1.8m^{-2}$. It is found that the flow rate under the view-limited condition decreases comparing with that under the normal view condition, and participants can learn from the experience.

We would like to mention that there are also some experimental studies on single file pedestrian flow. We do not review these studies, and the interested readers may refer to e.g., Seyfried et al. (2005), Chattaraj et al. (2009) and Yanagisawa et al. (2012).

2.2.2 Open corridor

Daamen and Hoogendoorn (2003) performed an experimental study on the effect of bottleneck in pedestrian flow. One narrow exit (width =1.0m) is used as the bottleneck for one large room. Approximately 80 pedestrians participate in the experiment. It turns out that only a small amount of the width is used at the location of the bottleneck. Further upstream, the pedestrian flow 'spreads out' and covers more or less the entire available width.

Isobe et al. (2004) performed an experimental study on bi-directional pedestrian flow in Japan. The width of the corridor is 2m and the length is 12m. Up to 70 participants are used. The pattern formation is observed in the experiment, while the jamming transition does not occur due to the finite size effect.

Moussaid et al. (2011) performed an experimental study on bi-directional pedestrian flow in France. The width of the corridor is 1.75m and the length is 7.88m. 40 participants are used in 20 runs. It is found that during the experiments, pedestrians apply two simple cognitive procedures to adapt their walking speeds and directions. Thus it is necessary to understand the crowd dynamics through cognitive heuristics.

Zhang et al. (2011, 2012) performed an experimental study on bi-directional pedestrian flow in Germany. The widths of the corridor are 3m and 3.6m in different runs, and the length is 8m. Up to 350 participants are used in 22 runs. The Voronoi method is used to study the velocity-density relationship, and the corresponding fundamental diagram is compared with some previous ones. It is found that the bi-directional flows are larger than the uni-directional ones when $\rho > 2m^{-2}$.

Plaue et al. (2011) performed an experimental study on intersecting flows at different situations in Germany. 46 participants are used. The widths of the corridor are set as 1.8m~3.6m in three different scenarios. The densities of the pedestrian flows in the video are measured by a nearest-neighbor kernel density method. Zhang and Seyfried (2014) compare the results in this paper with that in Zhang et al. (2011, 2012). It is found that intersecting angle has no influence on the fundamental diagram at least for 90° and 180°. The processes responsible for the reduction of the velocity or flow are independent from the intersecting angle.

Suma et al. (2012) performed an experimental study on the anticipation effect in bi-directional pedestrian flow. The width of the corridor is 3m and the length is 5m. 25 participants are used in three scenarios. The results show that the strength and range of anticipation significantly affect pedestrian dynamics, and there is one optimal strength of anticipation to realize the smoothest bi-directional flow.

Lian et al. (2015) performed an experimental study on four-directional intersecting pedestrian flows. The test track is two corridors which intersect vertically. The width of the two corridor is 3.2 m and the length is 11.2 m. In this experiment 364 participants are used, and the maximum local density can reach as high as $10m^{-2}$ in the cross area. However, the corresponding local flow rate is much larger than that observed in Hajj pilgrimage. This might because the cross area is very limited ($3.2*3.2m^2$), and it connects to two low-density corridors.

Feliciani and Nishinari (2016) performed an experimental study on the lane formation mechanism in bi-directional pedestrian flow. The widths of the corridor are 3m and the length is 10m. Five different phases are distinguished in the experiment, and the results show that balanced bidirectional flow is the most stable configuration after the lanes are formed. In this configuration, the lane formation process requires pedestrians to laterally move to a largest extent compared to low flow-ratio configurations.

To summarize, Table 1 lists the setups of the above mentioned experiments. We can find that except in Lian et al. (2015), the maximum densities are all smaller than $6m^{-2}$.

We would like to mention that there are many other open corridor experimental studies, see e.g., Helbing et al. (2005), Kretz et al. (2006), Wei et al. (2015), Jin et al. (2017). However, these papers focus on self-organized phenomena rather than the flow-density or velocity-density relationships. Although one does not know exactly the maximum density in these experiments, it is likely that the maximum density is below $6m^{-2}$, due to the similar experiment setup as that in Table 1.

Table 1. The setups of some open corridor experiments

| Experiment and time | Country | Properties | Maximum density ($m^{-2}$) |
|---|---|---|---|
| Daamen-Hoogendoorn (2003) | Netherlands | Bottleneck | 4.5 |
| Isobe et al. (2004) | Japan | Bi-directional | 2.9 |
| Moussaid et al. (2011) | France | Bi-directional | 2.3[*] |
| Zhang et al. (2011, 2012) | Germany | Bi-directional | 4.0 |
| Plaue et al. (2011) | Germany | Intersecting | 2.5 |
| Suma et al. (2012) | Japan | Bi-directional | 1.3 |
| Lian et al. (2015) | China | Intersecting | 10.0 |
| Feliciani-Nishinari (2016) | Japan | Bi-directional | 2.3 |

* The maximum occupancy is 0.8. The maximum density is calculated via 0.8*N/S. N =40 is the number of participants and S =$13.8m^2$ is the area of the corridor.

The literature review indicates that although there are many empirical and experimental studies of pedestrian flow, the data at high densities ($\rho > 6m^{-2}$) are still far from adequate. Therefore, it is necessary to carry out further studies on high density pedestrian flow.

**3. The experiment setup**

To study the pedestrian dynamics under high density situation, we organized the pedestrian flow experiment on Dec. 3th, 2016, in the Jiulonghu Campus of Southeast University of China. 278 students are recruited to take part in the experiment which lasts for about 2.5h. The experiment is organized on a large square, and plastic stools are used to form the boundaries, as shown in Fig.2. The radius of the inner circle is 2m, and that of the outer circle is 3.5m. So the area of the ring corridor is about $26m^2$. We use one unmanned aerial vehicle (UAV) to film the whole experiment. It hovers over the center of the two circles, and the height is about 11.8m. The video is 30 frames per second, and the resolution is 2704*1520. We use the software named *Tracker* (http://physlets.org/tracker/) to extract the data manually.

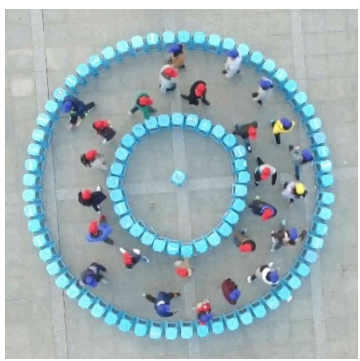

Fig.2. The basic configuration of the experiment.

Table 2. Details of each run in the experiment

| Run number | Actual Density ($m^{-2}$) | Unidirectional experiment | | | Bidirectional experiment | | |
|---|---|---|---|---|---|---|---|
| | | Number of participants | Walk direction | Duration of experiment (m:ss) | Number of participants walking clockwise | Number of participants walking anticlockwise | Duration of experiment (m:ss) |
| 9-1 | 9.04 | 235 | A | 7:40 | 117 | 118 | 9:40 |
| 8-1 | 7.69 | 200 | A | 4:28 | 101 | 99 | 5:05 |
| 7-1 | 6.27 | 163 | A | 4:39 | 78 | 85 | 4:11 |
| 6-1 | 5.65 | 147 | A | 2:32 | 70 | 77 | 4:06 |
| 5-1 | 4.85 | 126 | A | 2:00 | 65 | 61 | 2:03 |
| 9-2 | 8.35 | 217 | C | 3:51 | 106 | 111 | 2:21 |
| 8-2 | 7.65 | 199 | C | 2:31 | 100 | 99 | 3:23 |
| 7-2 | 6.42 | 167 | C | 2:19 | 84 | 83 | 4:10 |
| 6-2 | 5.69 | 148 | C | 1:50 | 74 | 74 | 2:09 |
| 5-2 | 4.73 | 123 | C | 1:48 | 59 | 64 | 2:10 |
| 4-1 | 4.04 | 105 | C | 1:29 | 53 | 52 | 1:39 |
| 3-1 | 3.19 | 83 | C | 1:06 | 41 | 42 | 1:08 |
| 2-1 | 2.15 | 56 | C | 1:07 | 28 | 28 | 1:02 |
| 1-1 | 1.00 | 26 | C | 1:01 | 13 | 13 | 0:53 |

In the experiment, we ask the participants to walk as usual at low densities and to move forward as far as they can at high densities. In each run, we first perform the uni-directional experiment. After some time, when the "steady state" is observed, we ask the participants to stop. Then we perform the bi-directional experiment by asking participants wearing blue and yellow[1] caps to turn around. When the "steady state" is observed again, we end this run of experiment. Table 2 shows the details of each run. The name of each run is defined as "predetermined density-order of run". For instance, "8-2" means the predetermined density is $8m^{-2}$ and the run is the second run. In the "walking direction", A means anticlockwise and C means clockwise. We would like to mention that the participant number and the actual density in some runs are slightly different from what we designed. This is because there are too many participants, and some participants did not follow the instruction. We have no time to count the participant number before each run of experiment.

**4. The experimental results of uni-directional flow**

This section discusses the experimental results of unidirectional pedestrian flow. Firstly, we present the fundamental diagram of system flow rate versus global density in Fig.3, in which each data point corresponds to one run of the experiment. Here the system flow rate is averaged over four equidistant locations. At each location, the flow rate is defined as Q/(WT). Here T is duration time of an experiment run. Q is the number of pedestrians that have crossed the location during the experiment run, and W is width of the test track (1.5m).

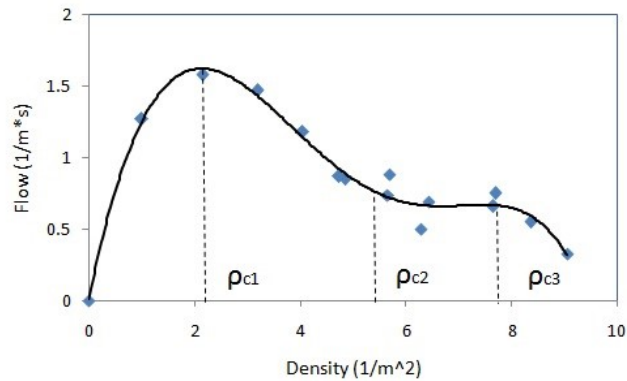

Fig.3. The fundamental diagram of uni-directional flow in the experiment. The solid line is the fitted line.

One can see that the there are three critical densities on the fundamental diagram. When the global density $\rho < \rho_{c1}$, the pedestrians are in ***free flow***. The flow rate increases with the density and reaches the maximum at $\rho = \rho_{c1}$. In the density range $\rho_{c1} < \rho < \rho_{c2}$, the pedestrian flow becomes ***congested***. The flow rate begins to decrease with the density. However, in the ***over-congested*** density range $\rho_{c2} < \rho < \rho_{c3}$, the flow rate becomes almost independent of the density. Finally, when the pedestrians become ***hyper-congested*** ($\rho > \rho_{c3}$), the flow rate decreases with the density again.

---
[1] In each run, one participant is randomly chosen to wear the yellow cap.

Now we investigate the evolution of flow rates in the congested, over-congested and hyper-congested states, see the three typical examples shown in Fig.4. Here the flow rate is calculated every 15 seconds and also averaged over four equidistant locations. Fig.5 shows the standard deviations of the time series of the flow rate of all the uni-directional runs. One can see that there exist three different density ranges. In the hyper-congested state, the fluctuation amplitude of the flow rate is quite large[2]. For example, in Run 9-1, the flow rate is very low for the first two minutes and becomes zero at around 2:45 minute. Then it abruptly increases to about 0.6 $m^{-1}s^{-1}$. The large flow rate is maintained for about two minutes, and then gradually decreases to almost zero again.

On the contrary, in the congested (e.g., Run 5-1) and over-congested states (e.g., Run 7-2), the fluctuation amplitude of flow rate is weaker and almost independent of the density[3]. In the free flow, the fluctuation amplitude of flow rate in free flow further decreases.

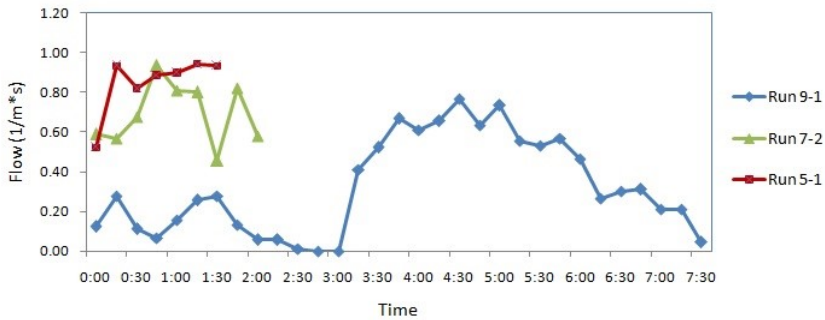

Fig.4. The flow rates in some uni-directional runs.

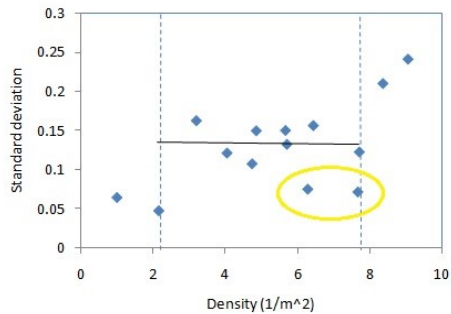

Fig.5. The standard deviations of the time series of the flow rates of all the uni-directional runs.

To further illustrate the fluctuation feature in different states, Fig.6 shows the cumulative angles travelled by the pedestrian wearing yellow cap in some typical runs. One can see that in the hyper-congested state (Fig.6(a)), the slope of the curve significantly changes during the experiment, which means the walking velocity significantly changes. On the other hand, in the over-congested (Fig.6(b)) and congested states (Fig.6(c)), the curve is nearly linear.

---

[2] If the density is so high that no pedestrian can move, then no fluctuation exists.
[3] The fluctuation amplitude of flow rate seems abnormally small in Run 7-1 and Run 8-2 (indicated by the circle in Fig.4), which might be due to the strong stochasticity in pedestrian flow. More experiments are needed to check this issue.

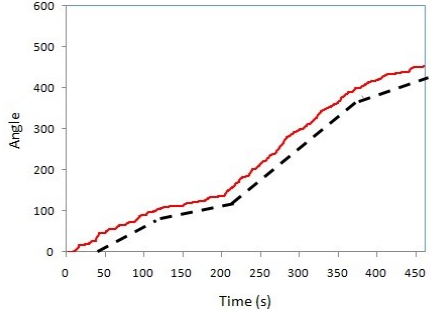

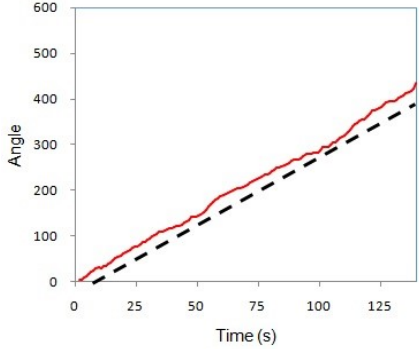 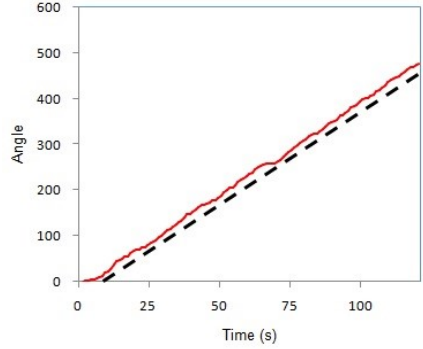

Fig.6. The cumulative angles travelled by the pedestrian wearing yellow cap. (a) Run 9-1; (b) Run 7-2. (c) Run 5-1. The dashed lines are guide for eyes.

Next we study evolution pattern of the pedestrian flow. To this end, we divide the test track into 8 equal subareas, see Fig.7. Fig.8 shows the spatiotemporal diagrams of local densities in some typical examples. Here the values are the pedestrian numbers in each subarea. One can see that the pedestrians' distribution is inhomogeneous, in particular at high densities. Moreover, the downstream propagation of pedestrian clusters can be observed at high densities, see Fig.8(a). The propagation velocity depends on the system flow rate, and it becomes almost zero when the system flow rate is small. With the decrease of the global density, the propagation process becomes not so manifest but still observable, see Fig.8(b). With the further decrease of the density, the propagation process vanishes, see Fig.8(c).

We would like to mention that emergence of the pedestrian clusters might be the reason why flow rate can keep nearly constant in the over-congested state. When the density is smaller than $\rho_{c2}$, it is very likely that the physical contacts of pedestrians are negligible. This might also be the reason why the density higher than $\rho_{c2}$ is seldom observed in many previous empirical data: people do not feel comfortable about the physical contacts. They would stop when the density approaches $\rho_{c2}$. If the density exceeds $\rho_{c2}$, some pedestrians have physical contacts with each other, and they form pedestrian clusters. Inside the clusters, pedestrians from behind push the ones in front. As a result, the flow rate could maintain a constant value until the density reaches $\rho_{c3}$.

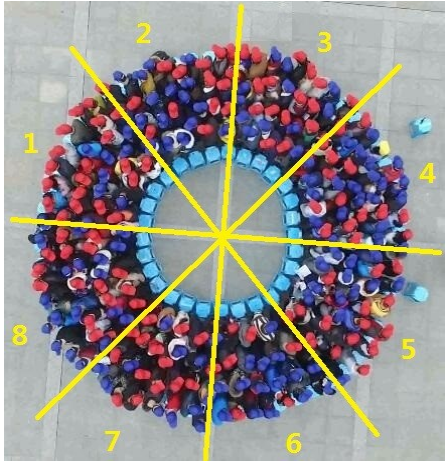

Fig.7. The division of 8 subareas. The snapshot is obtained from Run 9-1.

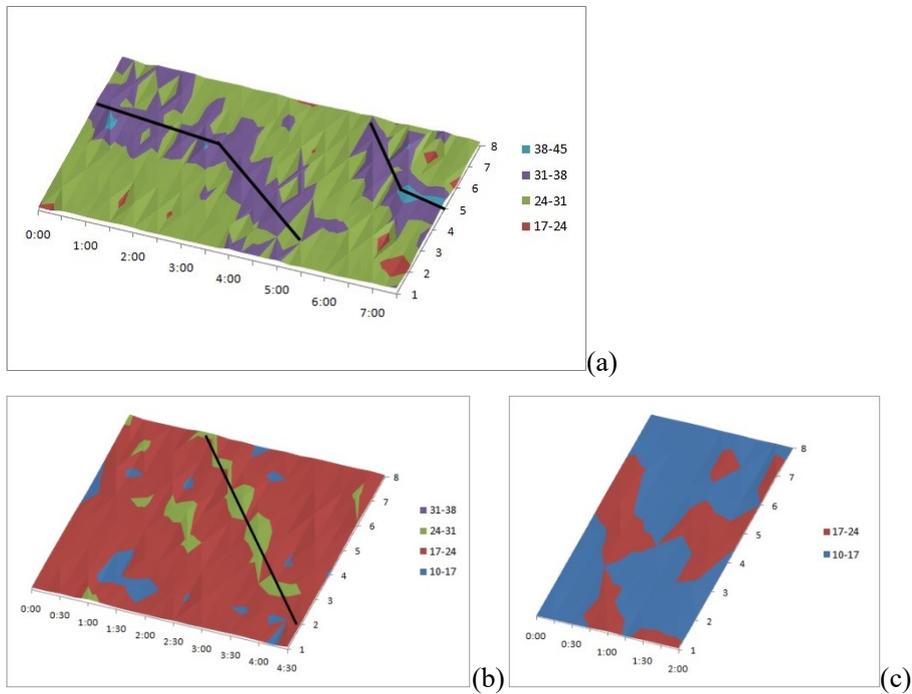

Fig.8. The spatiotemporal diagrams of local densities in some uni-directional runs. (a) Run 9-1; (b) Run 7-1. (c) Run 5-1. The black lines showing the propagation process are guide for eyes. Note the time scales are different in different subfigures.

Fig.9 shows the evolution of the standard deviations of the local densities in the eight subareas in several typical runs. Roughly speaking, in the hyper-congested state, the inhomogeneity changes significantly. For example, in Run 9-1, the system is quite homogeneous at T=1:45, at which the standard deviation is only 1.8. However, at T =7:00, the system becomes very inhomogeneous, at which the standard deviation reaches 7.5. In contrast, in the over-congested and congested states, the change amplitude of the inhomogeneity is relatively smaller.

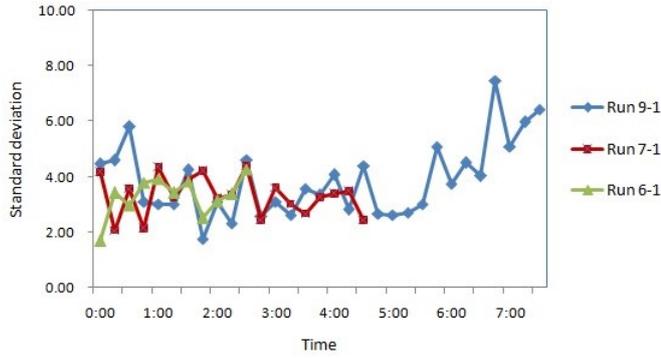

Fig.9. The standard deviations of the local densities in the 8 subareas in some uni-directional runs.

Fig.10 shows the average value of the time series of the standard deviations of the densities in all the runs, which can reflect the overall inhomogeneity of the system. One can see that two different density ranges can be classified. In the hyper-congested flow, the average value decreases with the density. In contrast, in the congested and over-congested flow, the average value increases with the density.

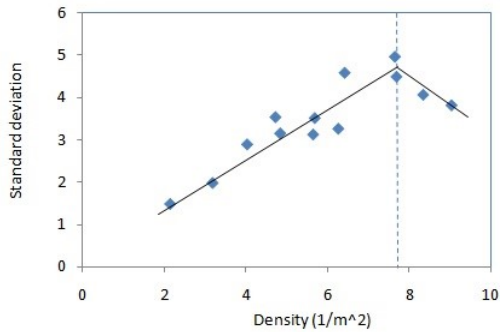

Fig.10. The average value of the time series of the standard deviations of the densities in all the uni-directional runs.

Finally, Fig.11 compares our experimental results with previous ones. One can see that our results are roughly in agreement with the empirical ones of Helbing et al (2007) in the free flow, over-congested and hyper-congested states. Helbing et al. (2007) pointed out that there could occur "crowd turbulence" at extremely high density. Our experiment shows that in the hyper-congested flow, the flow rate significantly fluctuates. Pedestrians sometimes transit between the stopped state and the moving state, which might correspond to the "crowd turbulence". On the other hand, the flow rate in the congested state is remarkably smaller in our experiment. This might be because we ask the participants to walk as usual at not so high density. In contrast, the pedestrians' walking behaviors in the empirical data are not normal in Helbing et al (2007): the data are obtained from the Hajj pilgrimage, and the pedestrians hope to walk as fast as possible.

In the data of Mori-Tsukaguchi (1987) and Hankin-Wright (1958), the flow rates in the congested state are also remarkably larger than that in our experiment. The former data are obtained from the commuters who usually walk fast. The latter data are obtained from one experiment. However, the instruction is not specified. A possible reason is that the participants were asked to move fast.

At very high densities, the data is lacked in Mori-Tsukaguchi (1987) and Weidmann (1993). The maximum density of these data roughly corresponds to $\rho_{c2}$ in our experiment. This might be because, as mentioned before, people do not feel comfortable about the physical contact. They would stop when the density approaches $\rho_{c2}$. Actually, if there is no instruction to move forward as far as they can, the participant in our experiment will not move under high-density situation.

Finally, the flow rate in Weidmann's data is much lower than that in our experiment. This might be because Weidmann's data are the averaged values of 25 datasets. In some datasets, the pedestrians might walk not fast, e.g., when they are shopping[4].

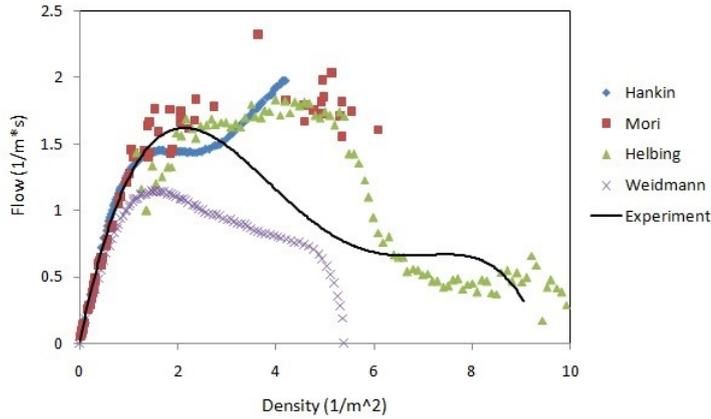

Fig.11. The comparison between our experimental results and previous uni-directional data.

## 5. The experimental results of bi-directional flow

Comparing with unidirectional pedestrian flow, one significant feature of bidirectional flow is the lane formation process. Therefore, we firstly discuss this process, then study the flow dynamics after the lane formation.

### 5.1. The lane formation process

This subsection studies the lane formation process. To this end, we propose two indices. They are, respectively, the average value (denoted by $\bar{r}$) and the standard deviation (denoted by $\sigma_r$) of the radial positions of the pedestrians walking in clockwise/anticlockwise direction.

Fig.12 shows three typical lane formation processes. Here C means the results for clockwise-moving pedestrians and A means that of Anticlockwise-moving ones. Fig.12(a) shows the first process (Run 9-1), in which $\sigma_r$ remarkably increases for one direction and decreases for the other direction. This corresponds to the situation that in the former direction, pedestrians split into inner lane and outer lane. In the latter direction, pedestrians gather into one middle lane. For the video of this lane formation process, see the Supplemental Video S1. Fig.13 shows one snapshot at T=0:43 in Run 9-1, in which the three lanes have formed. Notice that the pedestrian distribution is not homogeneous at this moment. A large group of pedestrians wearing red cap gathered together as indicated by the yellow circle.

---

[4] Unfortunately, the details of the 25 datasets are lacked in Weidmann's paper.

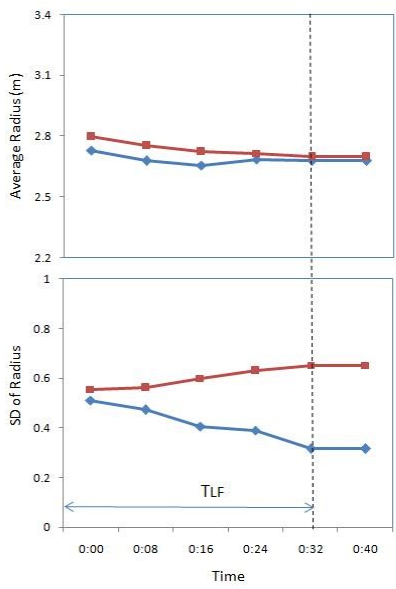
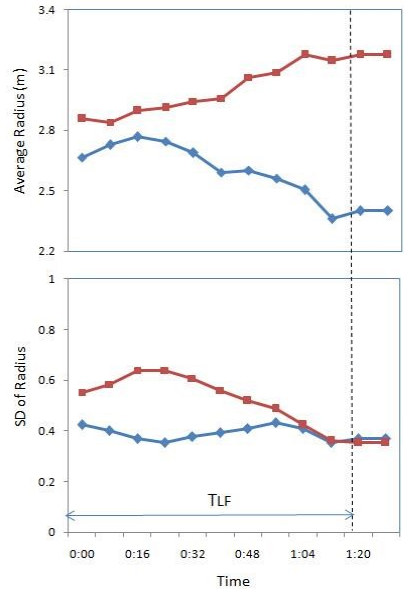
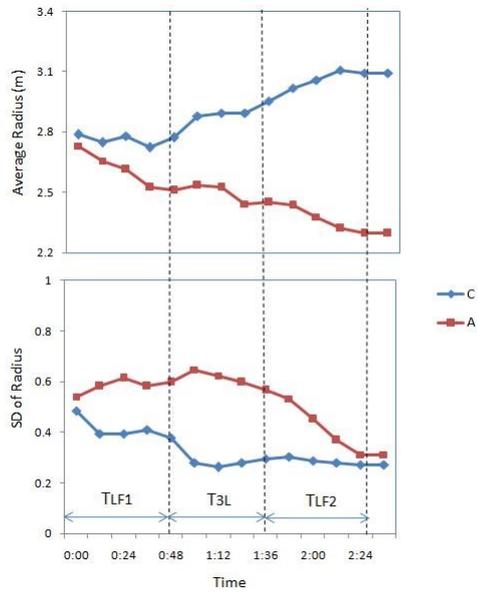

Fig.12. The evolution of pedestrians' radiuses during the lane formation process. (a) Run 9-1; (b) Run 8-2; (c) Run 8-1.

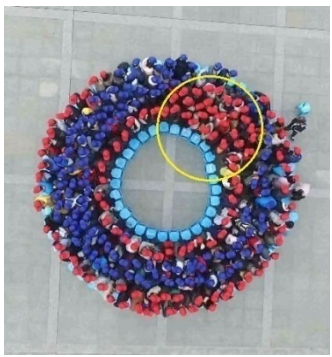

Fig.13. One video snapshot of the lane formation process in Run 9-1, T=0:43.

Fig.12(b) shows the second process (Run 8-2), in which $\bar{r}$ increases for one direction and decreases for the other direction. This is because in the former/latter direction, pedestrians gather into outer/inner lane. Although it might change remarkably in the lane formation process, $\sigma_r$ becomes almost equal for the two directions when the two lanes formed. For the video of this lane formation process, see the Supplemental Video S2. Fig.14 shows one snapshot at T=1:17 in Run 8-2, in which the two lanes have formed.

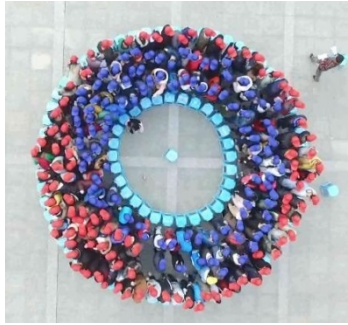

Fig.14. One video snapshot of the lane formation process in Run 8-2, T=1:17.

Fig.12(c) shows the third process (Run 8-1), in which the lane formation process can be classified into two sub-processes. In the first sub-process, three lanes have formed, similar to that shown in Fig.13. The three lanes are maintained for some interval, see the snapshot in Fig.15(a). At the beginning of the second sub-process, a gap appears in the middle lane. The leading pedestrian of outer lane crosses the middle lane through the gap (see the yellow circled pedestrian in the snapshot in Fig.15(b)), other pedestrians also follow the leader and cross. As a result, three lanes finally transit into two lanes (see the snapshot in Fig.15(c)). For the video of this lane formation process, see the Supplemental Video S3.

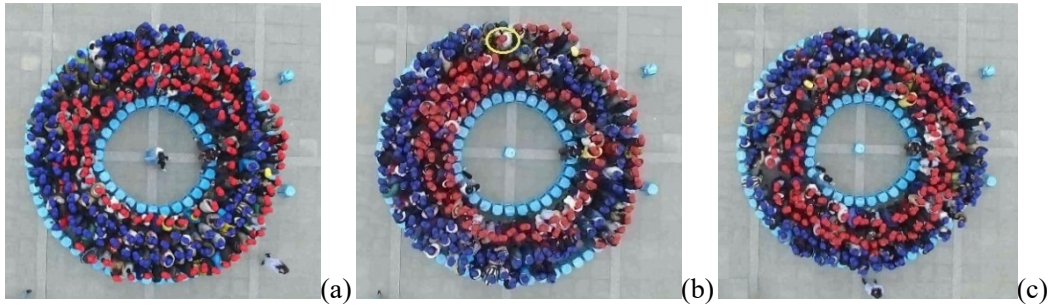

Fig.15. The video snapshots of the lane formation process in Run 8-1. (a) T=0:48; (b) T=1:36; (c) T=2:26.

For the 1st and 2nd types of processes, we study the lane formation time (denoted by $T_{LF}$). For the 3rd type of process, we study the lane formation time of the two sub-processes (denoted by $T_{LF1}$ and $T_{LF2}$, respectively) and the duration time of the three-lanes state ($T_{3L}$). Table 3 shows that the formation of three lanes is usually quick, even if at high densities. The longest time is about 48s in the 1st sub-process in Run 8-1. But at high densities, the direct formation of two lanes takes longer time (e.g., about 77s in Run 8-2). Moreover, the duration time of the three-lane state fluctuates significantly. For example, it lasts for 116 s in Run 7-2, but it is maintained only for 15 s in Run 6-2. We would like to mention that if the system width increases, it might form more than 3 lanes, which needs to be studied in future experiments.

Table 3. The lane formation processes of all the bi-directional runs

| Type of processes | Run number | $T_{LF}$ (s) | $T_{LF1}$ (s) | $T_{3L}$ (s) | $T_{LF2}$ (s) |
|---|---|---|---|---|---|
| 1 | 9-1 | 32 | | | / |
|   | 9-2 | 28 | | | |
|   | 6-1 | 16 | | | |
|   | 5-2 | 15 | | | |
| 2 | 8-2 | 77 | | | |
|   | 5-1 | 30 | | | |
|   | 4-1 | 11 | | | |
|   | 3-1 | 7 | | | |
|   | 2-1 | 7 | | | |
|   | 1-1 | 3 | | | |
| 3 | 8-1 | | 48 | 48 | 52 |
|   | 7-1 | | 30 | 82 | 16 |
|   | 7-2 | | 35 | 116 | 21 |
|   | 6-2 | | 20 | 15 | 15 |

**5.2. The flow dynamics after the lane formation**

This subsection studies the flow dynamics after the lanes have formed. Firstly, in Fig.16 we plot the fundamental diagram and compare with the uni-directional one. Based on the fitted lines, four density ranges are classified:

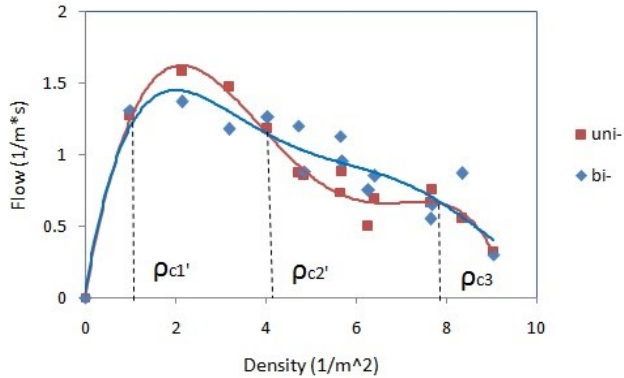

Fig.16. The uni-directional and bi-directional fundamental diagrams in the experiment.

(1) In the hyper-congested state ($\rho > \rho_{c3}$), the bi-directional flow rate is roughly equal to the uni-directional one. Fig.17 shows the evolution of 15-seconds bi-directional flow rates after the lane formation, and compares it with the uni-directional one. The significant fluctuation of flow rates is also observed. Fig.18 shows two video snapshots after the lane formation. It can be seen that the large group of pedestrians wearing red cap gradually disperse. Fig.19 shows the spatiotemporal diagrams of the local densities in Run 9-1. One can see the distribution of pedestrians is always inhomogeneous. In particular, no pedestrian cluster propagation has been observed.

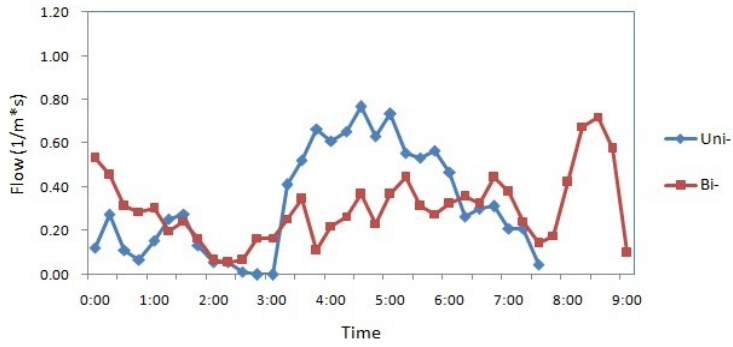

Fig.17. The flow rates of both uni- and bi-directional flow in Run 9-1.

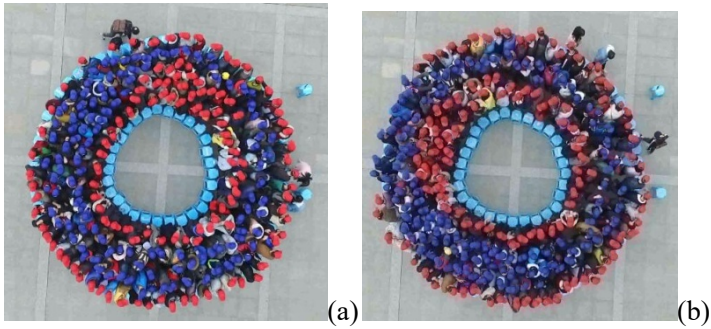

Fig.18. The video snapshots after the lane formation process in Run 9-1. (a) T=2:00; (b) T=6:45.

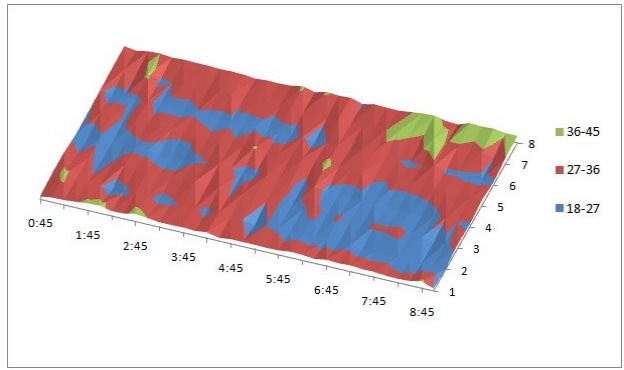

Fig.19. The spatiotemporal diagrams of local densities after the lane formation process in Run 9-1.

(2) In the density range $\rho_{c2'} < \rho < \rho_{c3}$, the bi-directional flow rate is larger than the unidirectional one. This is related to the inhomogeneous distributions of pedestrians in the formed lanes. Fig.20 shows one typical snapshot when three lanes are formed in Run 6-1. One can see that after lane formation, the pedestrian densities of the inner lane and the outer lane are significantly different. Although the density in the inner lane is small, pedestrians in the middle lane did not invade the space of inner lane. As a result, the flow rate of inner lane is about $0.88s^{-1}$, which is much large than that of the outer lane (about $0.22s^{-1}$). The high inner lane flow results in the larger averaged flow rate.

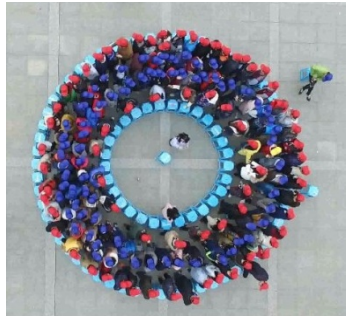

Fig.20. The video snapshot after the lane formation process in Run 6-1, T=1:09.

(3) In the density range $\rho_{c1'} < \rho < \rho_{c2'}$, the bi-directional flow rate becomes smaller than the unidirectional one. This is due to the formation of localized crowd. Fig.21(a) shows one typical snapshot in Run 4-1. One can see that at this moment, some pedestrians gathered together in the right part forming a localized crowd, in which they can only slowly move. In contrast, the density in the left part is very small, pedestrians can walk freely. The localized crowd dissipates after some time (see Fig.21(b)), and merges again (see Fig.21(c)). Fig.22 shows the spatiotemporal diagrams of the local densities in this run. One can see the formation and dissipation of localized crowd.

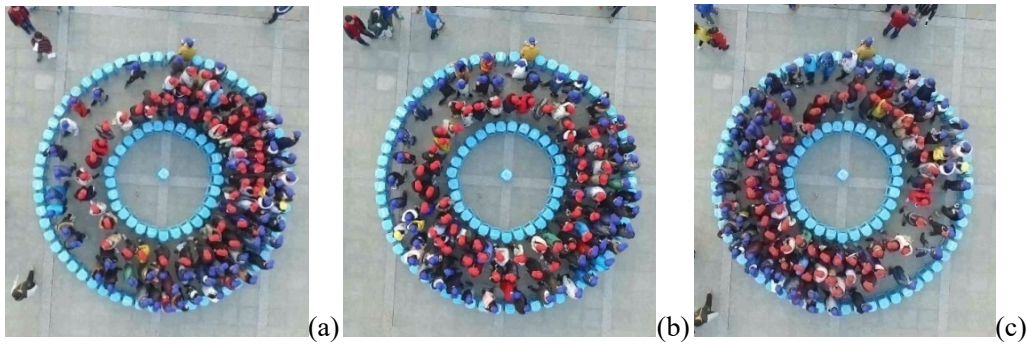

Fig.21. The video snapshots after the lane formation process in Run 4-1. (a) T=0:51; (b) T=1:13; (c) T=1:32.

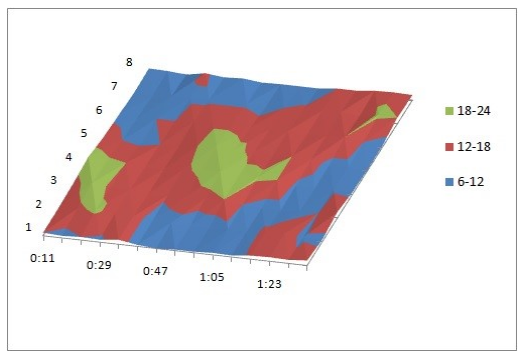

Fig.22. The spatiotemporal diagrams of the local densities after the lane formation process in Run 4-1.

(4) Finally, in the free flow ($\rho < \rho_{c1'}$), there is almost no interactions among pedestrians. Thus, the bi-directional flow rates are equal to the uni-directional one.

## 6. Conclusion

One main purpose of pedestrian dynamics study is to guarantee the safety of pedestrians in high density crowd, in which the stampede accident is prone to occur. However, the previous data at high densities ($\rho > 6m^{-2}$) are still far from enough. Motivated by the fact, we organized an experiment in a circular corridor, in which the maximum density reaches $\rho = 9m^{-2}$.

In the uni-directional flow experiment, four different states can be found, including the free flow, congested state, over-congested state and hyper-congested state. The features of the hyper-congested state are similar to the "crowd turbulence" reported in the empirical data of Helbing et al., in which the transition between the stopped state and the moving state can be observed. The downstream propagation of pedestrian clusters has been observed at high densities, which makes the flow rates in the over-congested state nearly constant.

In the bi-directional flow experiment, the lane formation processes can be classified into three different types: (1) three lanes are directly formed ; (2) two lanes are directly formed; (3) firstly three lanes are formed, and then they transit into two lanes. The averaged values of pedestrians' radiuses and their standard deviations have been used as the indicators for lane formation processes. After the lane formation, bi-directional flow rate exceeds unidirectional one's in the density range $\rho_{c2'} < \rho < \rho_{c3}$, due to the inhomogeneous distribution of pedestrians across the lanes. In contrast, bi-directional flow rate is smaller than unidirectional one's in the density range $\rho_{c1'} < \rho < \rho_{c2'}$, due to the formation of localized crowd.

Our experiment results are expected to be helpful for better understanding pedestrian dynamics and for pedestrian flow modeling at high densities. Finally, we would like to mention that our experimental data are still very limit. In the future work, more large-scale experiments involving more participants are needed.

## Acknowledgements

We are very grateful to Xue Jin, Zi-Xin Wu, Jin-Lian Wu, Jun-Lan Chen, Wei Wei, Ming-Zhe Zang, Jun-Lin Yin, Yang Zhao, Ling-Yu Meng, Xiao-Bo Dong for their help to extract the data from the experimental video. This work was funded by the National Natural Science Foundation of China (No. 11302022, 11422221, 71371175, 71621001), and the Natural Science Foundation of Jiangsu Province (No. BK20150613, BK20150619).